\documentclass{optica-article}

\journal{opticajournal} 

\articletype{Research Article}

\usepackage{multicol}
\usepackage{lineno}
\UseRawInputEncoding
\begin{document}
	
	\title{Twist Bilayer Photonic slab's Angle-Dependent Guided Resonance Analysis based on Multiple Scattering}
	
	\author{Wenzhu Xie\authormark{1,2}, Yan Wang\authormark{2,6},Jingxuan Chen\authormark{1}, Jiahao Si\authormark{1,3}, Wei Rao\authormark{6,*}, Mingjin Wang\authormark{1,4,5,*}, Wanhua Zheng\authormark{1,2,3,4,5,*}}
	
	\address{
		
		\authormark{1}Laboratory of Solid State Optoelectronics Information Technology, Institute of Semiconductor, Chinese Academy of Sciences, Beijing, 100083,China\\
		\authormark{2}College of Future Technology, University of Chinese Academy of Sciences, Beijing, 101408, China\\
		\authormark{3}Center of Materials Science and Optoelectronics Engineering, University of Chinese Academy of Sciences, Street, Beijing, 100083, China\\
		\authormark{4}State Key Laboratory on Integrated Optoelectronics, Institute of Semiconductor, Chinese Academy of Sciences, Beijing, 100083, China\\
		\authormark{5}Hangzhou Institute for Advanced Study, University of Chinese Academy of Sciences, Hangzhou, 310024, China\\

		\authormark{6}Beijing Key Laboratory of CryoBiomedical Engineering and Key Laboratory of Cryogenics, Technical Institute of Physics and Chemistry, Chinese Academy of Science, Beijing, 100190, China}
	
	\email{\authormark{*}weirao@mail.ipc.ac.cn, mingjinwang@semi.ac.cn, whzheng@red.semi.ac.cn} 
	
	
	\begin{abstract*} 
		We present an analysis of the transmission spectra of the twisted bilayer photonic slabs using a modified rigorous coupled wave (RCWA) analysis, where the evanescent bases are replaced by bases with non-zero flux density. By utilizing the modified RCWA we demonstrate the calculation of eigenmodes, which has not been realized before. To counter for the transmission property, we propose a five-layer uniform slab approximation, with an accuracy around 0.04a/c, which is more straightforward and accessible for optical engineers compared to work by Lou \textit{et al}. [Phys. Rev. Lett. 126, 136101]. The moir\'{e} pattern perturbation induces a split of resonance, which show great potential for engineering the band structure. Moreover, We observe two distinct transmission phases: the angle-dependent phase and Fabry-P\'{e}rot phase, which is explained by a coupled-mode theory (CMT) with expanded channels brought by the modified eigenmodes. Our work provides a theoretical framework for the design and optimization of twisted bilayer photonic devices.
		
	\end{abstract*}
	

	\section{Introduction}\label{sec1}
	
	In recent years, there has been a heated discussion about moir\'{e} physics, which involves one layer stacked on top of an identical one with a relative in-plane twist angle. This moir\'{e} 2D material has a lot of interesting properties compared to conventional structure. The striking phenomenon of hybridization induced flat band leads to two insulating states--bandgap induced insulating state\cite{cao_superlattice-induced_2016,fang_electronic_2016} and unconventional correlated insulating state\cite{cao_correlated_2018,han_correlated_2024}. Doping away from correlated insulating state, superconductivity emerges\cite{cao_unconventional_2018,arora_superconductivity_2020,stepanov_untying_2020,guo_superconductivity_2025}, which severs as a fascinating platform to investigate the strongly correlated system. Meanwhile, heartened by the intriguing effect in moir\'{e} physics\cite{cao_superlattice-induced_2016,fang_electronic_2016,cao_correlated_2018,han_correlated_2024,cao_unconventional_2018,arora_superconductivity_2020,stepanov_untying_2020,guo_superconductivity_2025,kim_electrostatic_2024,tanaka_superfluid_2025}, there has seen a surge of research on the electronic counterpart twisted bilayer photonic crystal. Similarly, flat band is also observed in both 1D and 2D photonic structures\cite{dong_flat_2021,tang_modeling_2021,wang_intrinsic_2022,yi_strong_2022,hoang_photonic_2024,nguyen_magic_2022,huang_moire_2022}, specifically bound states in the continuum (BICs)\cite{huang_moire_2022,qin_arbitrarily_2023,qin_optical_2024} is also realized in moir\'{e} photonic structure. Moreover, the flat band indicates the localization behavior\cite{wang_localization_2020,zeng_localization--delocalization_2021,xu_hydrodynamic_2024} of the optical field. Magic angle laser\cite{mao_magic-angle_2021,raun_gan_2023,luan_reconfigurable_2023}, LiDAR beam steering module\cite{lou_free-space_2024} and angle-tunable filter\cite{lou_tunable_2022-1} are also demonstrated.
	
	On the other hand, the photonic crystal slab as one part of the twisted bilayer slab is the key optical modulation element of the photonic crystal surface emitting laser (PCSEL). The periodic lattice modulation of the photonic crystal slab enable a high quality factor (>$10^6$), a sub-degree beam divergence and single-mode stability under high-power operation (>10W)\cite{inoue_general_2022,yoshida_high-brightness_2023}. The photonic surface emitting laser can also be designed to lase at a specific direction serving as an active beam steering module\cite{sakata_dually_2020,wang_active_2024}. The twisted bilayer photonic structure incorporating the interlayer degree of freedom is more powerful in tailoring the optical field and thus is a promising platform for the next generation LiDAR optical steering module\cite{lou_free-space_2024}, laser\cite{mao_magic-angle_2021,raun_gan_2023} and reconfigurable devices\cite{luan_reconfigurable_2023}.  
	
	However, the understanding of the resonance of the twisted bilayer structure remains incomplete. (i) Calculation of the resonance mode of the twisted bilayer structure remains a difficulty. The commonly used supercell approximation\cite{tang_modeling_2021,wang_intrinsic_2022,tang_experimental_2023} neglects the effect of higher order moir\'{e} wave vectors, while the RCWA\cite{lou_theory_2021,lou_tunable_2022,lou_rcwa4d_2025} is insufficient to compute the eigenmodes of the system. Additionally, as the frequency increases, the resonance initially exhibits Fabry-P\'{e}rot characteristics, followed by angle-dependent feature. (ii) No study have ever reveled the transition mechanism of angle-dependent resonance from the Fabry-P\'{e}rot resonance. It is hard to determine the critical transition frequency between these two phases. This may relate to the scattering strength of fundamental reciprocal lattice vectors in individual slabs, but scattering becomes significant only within the narrow resonance bandwidth of an individual slab\cite{fan_analysis_2002}. However, this contradicts the broad frequency range of angle-dependent resonance.
	
	In this work, we propose a modified rigorous coupled wave analysis (RCWA)\cite{li_new_1997}, which is capable of calculating the eigenmodes of the twisted bilayer structure and the critical transition frequency between the angle-dependent regime and the Fabry-P\'{e}rot regime. In this modified framework, the evanescent bases are replaced by bases with non-zero energy flux density to appropriately describe the multi-reflection physics within the air gap of the bilayer system. An expanded basis set is also adopted to properly describe the quasi-periodic physics\cite{lou_theory_2021,lou_rcwa4d_2025}. By utilizing the scattering matrix of the two slabs, the eigenmodes of the system can be determined by solving an eigenvalue problem. We first propose a five-layer uniform slab mode approximation with an accuracy around 0.04$a/c$, whose validity is further confirmed by the eigenmode analysis. This approximation is more straightforward and accessible to optical engineers compared to the result of Lou \textit{et al}.\cite{lou_theory_2021}. Moreover, a group theory analysis is conducted to counter for the resonance split, which shows the interlayer modulation effect on the mode of the optical structure. Remarkably, a coupled mode analysis\cite{fan_analysis_2002,manolatou_coupling_1999,fan_temporal_2003} shows that the critical transition frequency is determined by the resonance of the photonic crystal slab mode under the whole passages including the expanded passages brought by the modified eigenmodes in contrast to the typical coupled mode analysis. Our study pave the wave for the twisted bilayer structure based photonic devices.
	\section{Method}\label{sec2}
	\subsection{Theoretical approach -- Modify RCWA}\label{sec2sec1}
\begin{figure}[h]
		\centering
		\begin{minipage}{1\linewidth}
			\centering
			\includegraphics[width=0.8\linewidth]{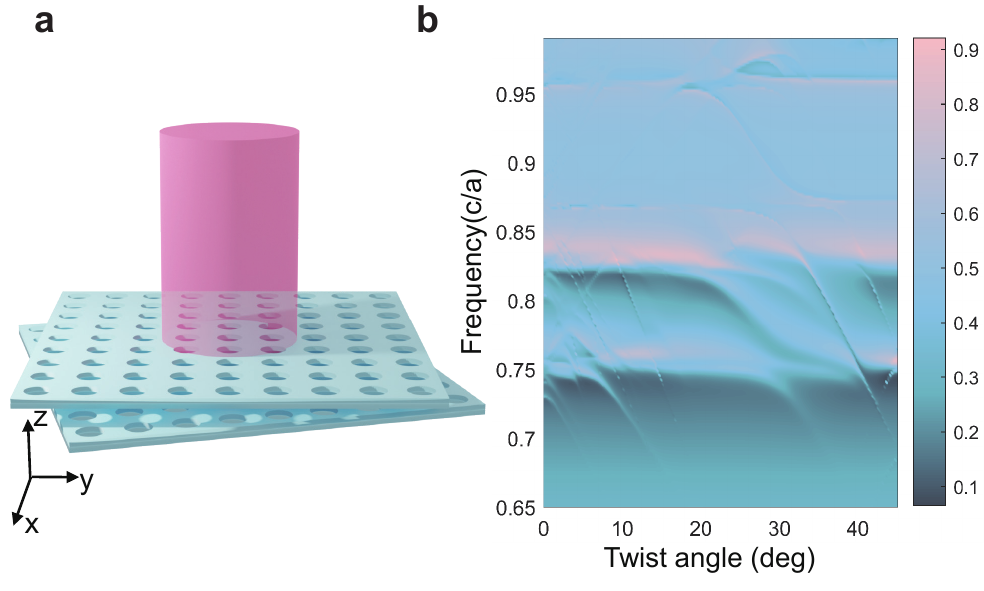}
			
		\end{minipage}
		\caption{(a) Twisted bilayer photonic crystal slabs illuminated by normally incident light polarized along the x-axis. Two stacked slabs are identical, with a relative in-plane twist angle, each with circular air holes. $\varepsilon_r=4,\ h=0.2a, \ d=0.3a, \ r=0.3a$, where $a$ is the period, $h$ is the thickness of the slab, $d$ is the thickness of the air gap between the two slabs and $r$ is the radius of the air hole.  (b) Angle-dependent guided resonance of the system in (a). The transmission coefficient is defined by the ratio of transmitted energy flux to incident flux perpendicular to the slab surface.}\label{fig1}
	\end{figure}

	The system under investigation is schematically depicted in Fig.\hyperref[fig1]{1(a)}. For simplicity, the slab's thickness is regulated to be small enough ($h<a/(2\sqrt{\bar{\varepsilon}-1})$) to ensure the respective uniform slab is in the single mode regime, $h=0.2a$. $\varepsilon_r=4 ,\ d=0.3a ,\ r=0.3a$, where $a$ is the period,  $d$ is the thickness of the air gap and $r$ is the radius of the air hole. 
	
	The twisted bilayer photonic crystal slabs exhibit intriguing transmission patterns in Fig.\hyperref[fig1]{1(b)}, which arise from the interlayer coupling. As the angle changes, the band structure varies due to the change in interlayer interaction and the spectral response exhibits dramatic angular dependence. An intuitive understanding of the system's physical mechanism is that the successively scattering processes by reciprocal lattice vectors of slabs 1 and 2 govern its optical response. To calculate the scattering matrix appropriate for the multiple scattering description, a modification is necessary to replace the evanescent bases in RCWA calculation by waves with non-zero energy flux density (more detailed information is in Supplemental Material, Sec.1 and Sec.2).
	
	With the modified new bases, we then start to formulate the scattering matrices. Each layer can scatter the incident light with in-plane wave vector $\boldsymbol{k}_{\text{inc}}$ to $\boldsymbol{k}_{\text{inc}}+\boldsymbol{G}^{(i)}$, where $\boldsymbol{G}^{(i)}$ is the reciprocal lattice vector of the $i$-th slab and $\boldsymbol{k}_{\text{inc}}$ is the in-plane wave vector of the incident light. The reciprocal space of the two slabs: $\mathcal{L}_1\equiv\{(m\hat{i}+n\hat{j})2\pi/a \textbar m,n\in\mathbb{Z}\}\equiv\{\boldsymbol{G}_{m,n}^{(1)}\textbar m,n\in\mathbb{Z}\}$, $\mathcal{L}_2\equiv\{(m\hat{i^{'}}+n\hat{j^{'}})2\pi/a \textbar m,n\in\mathbb{Z}\}\equiv\{\boldsymbol{G}_{m,n}^{(2)}\textbar m,n\in\mathbb{Z}\}$, with $\hat{i}$ and $\hat{j}$ denoting the unit vectors along the positive x-axis and y-axis respectively, $\hat{i^{'}}=cos(\theta)\hat{i}+sin(\theta)\hat{j}$ and $\hat{j^{'}}=-sin(\theta)\hat{i}+cos(\theta)\hat{j}$, where $\theta$ is the twist angle. Denote $\mathcal{L}=\mathcal{L}_1\oplus\mathcal{L}_2$. $\mathcal{L}$ represents the direct sum of  the reciprocal spaces of slab 1 and slab 2. $\mathcal{L}+\boldsymbol{k}_{\text{inc}}$ includes all the in-plane wave vectors involved in the scattering process of the two slabs.
	
\begin{figure}[h]
		\centering
		\includegraphics[width=0.6\textwidth]{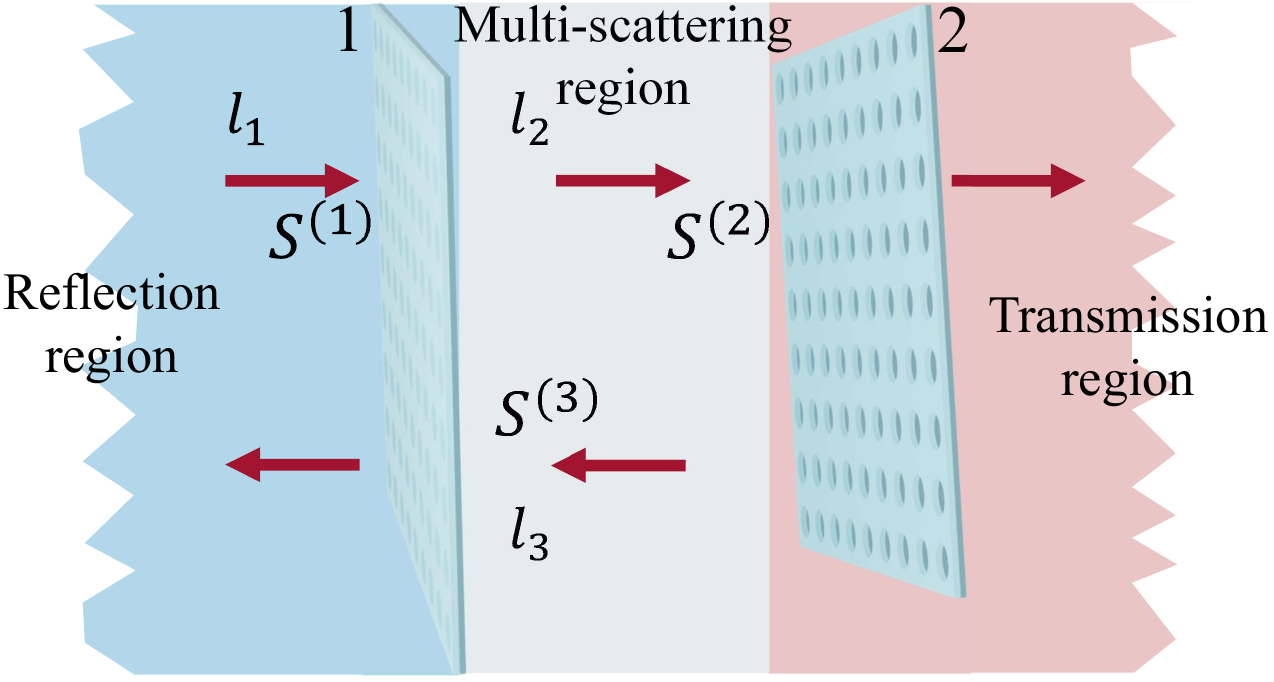}
		\caption{The physics that the three scattering matrices represent. $S^{(1)}$ quantifies light scattering along path $l_{1}$ caused by slab 1. $S^{(2)}$ governs light scattering through path $l_{2}$ originating from slab 2. $S^{(3)}$ defines light scattering along path $l_{3}$ due to slab 1. }\label{fig2}
	\end{figure}

	Our theoretical approach comprises three scattering matrices derived via the modified RCWA. RCWA is a semi-analytical Fourier-domain technique that offers high computational efficiency for analyzing electromagnetic fields in periodic systems\cite{li_new_1997,rumpf_improved_2011}. The scattering process is described by 
	\begin{equation} \begin{split}
			\left[
			\begin{array}{c}
				\boldsymbol{d}^{(1)} \\
				\boldsymbol{u}^{(2)} 
			\end{array}
			\right]
			=\boldsymbol{S}^{(i)} 
			\left[
			\begin{array}{c}
				\boldsymbol{u}^{(1)} \\
				\boldsymbol{d}^{(2)} 
			\end{array}
			\right].
	\end{split} \end{equation}\label{eq1}  
	$i=1,2,3$ stand for the three scattering matrices, which represents three physical processes in Fig.\hyperref[fig2]{2}. The character $u$ stands for upward and $d$ for downward. In this case, $\boldsymbol{d}^{(2)}$ is a zero vector, and only the left submatrix of  $\boldsymbol{S}^{(i)}$ contributes. The vector $\boldsymbol{u}^{(2)}$ or $\boldsymbol{d}^{(2)}$ includes the amplitude of free space eigenmodes as labeled by the in-plane wave vectors in $\mathcal{L}+\boldsymbol{k}_{\text{inc}}$ and polarization. However, standard RCWA constructs scattering matrices involved with solely modes with in-plane wave vector in $\mathcal{L}_1+\boldsymbol{k}_{\text{inc}}$ or $\mathcal{L}_2+\boldsymbol{k}_{\text{inc}}$. To cover the modes labeled by both slabs' reciprocal spaces, one can sweep $\boldsymbol{k}_{\text{inc}}$ across the reciprocal lattice space of either slab and synthesize the scattering matrices into a comprehensive one\cite{lou_theory_2021,lou_rcwa4d_2025}. There are no specific rules for the arrangement of various eigenmodes, but only to maintain consistency in both sides of equation \hyperref[eq1]{(1)}. Note that it is important to obey the natural boundary condition outside of the two slabs in order to align with the electromagnetic boundary condition of the original two-slab system. As a consequence, only free space eigenmodes within the inner air gap can be replaced by the modified eigenmode set (more detailed information is in Supplemental Material, Sec. 3).
	
	By employing the scattering matrices, the transmission field can be expressed as  
	\begin{equation} \begin{split}
			\boldsymbol{u}^{(3)}=\sum_{i=0}^{\infty} \boldsymbol{S}^{(1)}_{11}(\boldsymbol{S}^{(2)}_{21}\boldsymbol{S}^{(3)}_{21})^{i}\boldsymbol{S}^{(2)}_{21}\boldsymbol{u}^{(1)}.
	\end{split} \end{equation}\label{eq2}
	In this case, the eigenfunction of matrix $\boldsymbol{S}^{(2)}_{21}\boldsymbol{S}^{(1)}_{21}$ dictate the stationary electromagnetic field distribution (eigenfield) localized between the slabs. Furthermore, the eigenfield with an eigenvalue $\lambda$ that approximates 1 can significantly contribute to the characteristic resonance shape due to the multi-interference process among the scattered fields.
	\subsection{Experimental approach}\label{sec2sec2}
	The experimental structure consists of twisted bilayer photonic slabs with circular air hole. The slab is made of poly(ether-ether-ketone) (PEEK), with a relative dielectric constant of 3.1 as determined by time domain spectroscopy technique(TDS) utilizing a disk made of PEEK powder. The PEEK photonic crystal slab is supported by a 3D-printed scaffold made of a photosensitive resin material named Goda\textregistered8111X. The radius of the circular hole is R = 1.02mm, with a periodicity of 3.4mm. The spacing between the two layers of the photonic crystal slabs is 1.02mm, the thickness of each layer is 1.02mm as well. With this setup, the twist angle between the two circular hole arrays can be continuously varied from $0^{\circ}$to $360^{\circ}$. To evaluate the performance of the twist-angle bilayer slabs, a Ceyear AV3672D Vector Network Analyzer (VNA) and two Frequency Extension Modules (FEMs) are employed for measurements, with an operating frequency range from 75GHz to 110GHz. During the transmission spectrum measurement, one photonic slab was fixed while the other was rotated from $0^{\circ}$to $45^{\circ}$.
	\section{Result}\label{sec3}
	\subsection{Five-layer uniform slab approximation}\label{sec3sec1}

\begin{figure}[h]
		\centering
		\begin{minipage}{1\linewidth}
			\centering
			\includegraphics[width=0.8\linewidth]{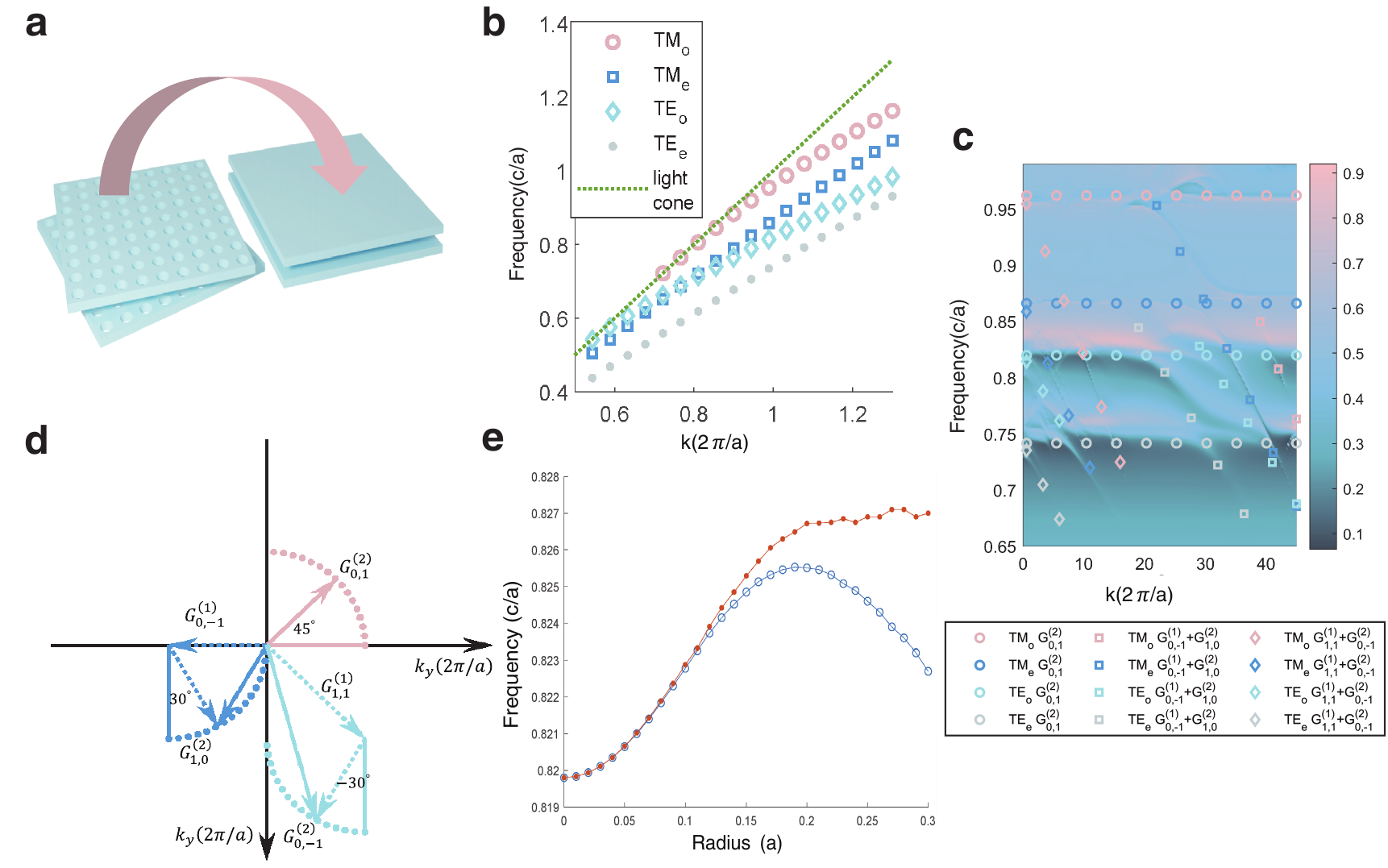}  		
		\end{minipage}
		\caption{(a) The corresponding five-layer uniform slab is surrounded by air, with the dielectric constant of the middle two slabs being $\bar{\varepsilon}$. (b) The $\omega-k$ dispersion relations of the five-layer slab are plotted. The suffix $o$ denotes the odd mode, and $e$ denotes the even mode. Here, $k$ is the in-plane wave vector. (c) The resonance frequencies are provided via $\omega_{i}(\boldsymbol{k}_{\text{inc}}+\boldsymbol{G}_{m_1,n_1}^{(1)}+\boldsymbol{G}_{m_2,n_2}^{(2)})$ ($i=1-4$), where $\omega_{i}$ corresponds to the dispersion relations in Fig.3(b),  with 1 for $TM_o$, 2 for $TM_e$, 3 for $TE_o$, and 4 for $TE_e$. (d) The examples of the evolution of the in-plane moir\'{e} vector, in the cases of  $\boldsymbol{G}_{0,1}^{(2)}$ for $\theta=45^{\circ}$, $\boldsymbol{G}_{0,-1}^{(1)}+\boldsymbol{G}_{1,0}^{(2)}$ for $\theta=30^{\circ}$and $\boldsymbol{G}_{1,1}^{(1)}+\boldsymbol{G}_{0,-1}^{(2)}$ for $\theta=-30^{\circ}$ are demonstrated. The one-fourth circle dashed lines indicate the trajectory of the moir\'{e} vector with the origin being the star point. The dashed lines with arrows represent the reciprocal lattice vectors of the two slabs, while the solid lines represent the moir\'{e} vectors. (e) Frequency of the two doubly degenerate E modes as a function of radius $r$ of the hole in both slabs. For each radius the average dielectric constant for both slabs is fixed to the value for $r=0.3a$.}\label{five-layer}
	\end{figure}

	Indeed, the twist-angle-dependent modes and the twist-angle-independent modes can be well explained by the modes of the corresponding five-layer uniform slab waveguide in Fig.\hyperref[five-layer]{3(a)}. This is different from the previous theoretical approach using the three-layer uniform slab modes\cite{lou_tunable_2022,lou_theory_2021}, which attempts to cover the transmission characteristic by solely one dispersion relationship. The system is governed by the master function below\cite{joannopoulos_photonic_2008}:
	\begin{equation} \begin{split}
			\left(\hat{H}+\hat{H}^{'}\right) \left|\phi \right\rangle=\frac{\omega^{2}}{c^{2}}\left|\phi \right\rangle,
	\end{split} \end{equation}\label{masterfunction}
	where $\hat{H}=\nabla \times \frac{1}{\bar{\varepsilon}}\nabla \times$ and  $\hat{H}^{'}=-\nabla\times\frac{\Delta\varepsilon}{(\bar{\varepsilon}+\Delta\varepsilon)\bar{\varepsilon}}\nabla\times$, $\bar{\varepsilon}$ represents the average dielectric constant, while $\Delta\varepsilon$ represents the fluctuating part. The ket vector $\left|\phi \right\rangle$ denotes the magnetic field distribution. We denote $V=\bigcup_{i=1}^{4} f_{i}(\mathcal{L}+\boldsymbol{k_{\text{inc}}})$, where $f_{i}$ associates each element of $\mathcal{L}$ with a five-layer uniform slab eigenmode characterized by the corresponding in-plane wave vector. The index $i$ specifies the polarization state and parity symmetry of the eigenmode. From a perturbation perspective, the approximate eigenfield can be expressed as $\Phi=\sum_{\phi_{i}\in V}a_i\phi_{i}$. The five-layer slab modes are the eigenmodes when all fluctuations are neglected, which is a good approximation since the coupling strength  $\left\langle\phi_2 \right|\hat{H}^{'}\left|\phi_1 \right\rangle$ becomes non-negligible only proximate to intersection of the unperturbed transmission resonance described by the dispersion relation of Fig.\hyperref[five-layer]{3(b)}, where the no-level crossing phenomenon becomes dominant\cite{sakurai_modern_2021}. The interaction is smaller with a larger distance between the two in-plane wave vectors and eigenvalues of the respective unperturbed eigenmodes. The five-layer uniform slab precisely predict the resonance of the system in a series of parameter with an accuracy around 0.04a/c. Note that the avoided crossing is not expected to be found near the intersection of TE modes and TM modes, because they are orthogonal with respect to the inner product defined by $\hat{H}^{'}$.

	The two slabs can cooperatively scatter light, converting the in-plane wave vector from $\boldsymbol{k}_{\text{inc}}$ into $\boldsymbol{k}_{\text{inc}}+\boldsymbol{G}_{m_1,n_1}^{(1)}+\boldsymbol{G}_{m_2,n_2}^{(2)}$, thereby excite the waveguide modes leading to guided resonance. This occurs when both the frequency and in-plane wave vector are approximately matched\cite{fan_analysis_2002} according to the  $\omega$-$k$ dispersion relation in Fig.\hyperref[five-layer]{3(b)}. As the angle varies, the resonant frequency changes corresponding to $\omega_{i}(\boldsymbol{k}_{\text{inc}}+\boldsymbol{G}_{m_1,n_1}^{(1)}+\boldsymbol{G}_{m_2,n_2}^{(2)})$ ($i=1-4$), resulting in the characteristic angle-dependent transmission features. The index $i$ denotes four dispersion relationships in the $\omega-k$ graph, with 1 for $TM_o$, 2 for $TM_e$, 3 for $TE_o$, and 4 for $TE_e$. The letters $o$ and $e$ represent odd and even modes with respect to the middle xy plane of the air gap. The approximation using the five-layer uniform slab eigenmodes accurately predicts the resonance in Fig.\hyperref[fig1]{1(b)} and Fig.\hyperref[five-layer]{3(c)} demonstrating excellent agreement between the simulated resonances and five-layer model predictions, indicating direct mode correspondence.
	
	The four twist-angle-independent guided resonances originate from $\omega_{i}(\boldsymbol{k}_{\text{inc}}+\boldsymbol{G}_{m_1,n_1}^{(j)})$ ($i=1-4$; $j=1,2$), in the case that the magnitude of the wave vector is invariant with the twist angle varied ($\boldsymbol{k}_{\text{inc}}=\boldsymbol{0}$). The $TM_o$ is the resonance with the highest frequency and is located around 0.95$c/a$, 0.87 $c/a$ for $TM_e$, 0.82 $c/a$ for $TE_o$, and 0.75 $c/a$ for $TE_e$. All four kinds of resonance relate to a moir\'{e} vector magnitude of $2\pi/a$.
	
	As for the twist-angle-dependent guided resonance, when the angle varies, the magnitude of the moir\'{e} vector changes, the orange and yellow trajectories in Fig.\hyperref[five-layer]{3(d)} for example, rendering the change in resonant frequency. The expected avoided crossing occurs at the intersection of modes of the same polarization, whereas the crossing emerges when modes with different polarizations intersect. The crossing is exemplified by the intersection of the blue circle and the gray parallelogram in Fig.\hyperref[five-layer]{3(c)}.
	
	The resonance split in transmission spectra can be understood in the perspective of group theory and perturbation. The split phenomenon is observed in Fig.\hyperref[fig1]{1(b)}, Fig.\hyperref[five-layer]{3(c)}, especially for the twist-angle-independent resonance with a splitting magnitude about 0.005$c/a$. Due to the effect of the moir\'{e}-pattern holes in both slabs, twisted bilayer system at an incommensurate angle exhibits a $D_{\text{4}}$ symmetry. Thus, the modes can be classified by $D_{\text{4}}$ point group through its four one-dimensional irreducible representations $A_1, A_2, B_1, B_2$ and one two-dimensional irreducible representation $E$, where only $E$ modes can be stimulated by the incident light. There are two different kinds of $E$ modes, which indicates that the resonance mode can split into two doubly degenerate modes that contribute to the transmission process(see supplemental). The mode splitting is demonstrated in \hyperref[five-layer]{Fig.4(e)}, where we plot the split of resonance near 0.82$c/a$ as a function of both slab's radius. It is note worthy that by utilizing the modulation of the moir\'{e} pattern can we manipulate the mode at $\Gamma$ point to realize high performance photonic crystal laser. Even if the laser usually works in the singly degenerate mode at $\Gamma$ point.

	\subsection{eigenmode analysis}\label{sec3sec2}	
\begin{figure}[h]
		\centering
		\includegraphics[width=1\textwidth]{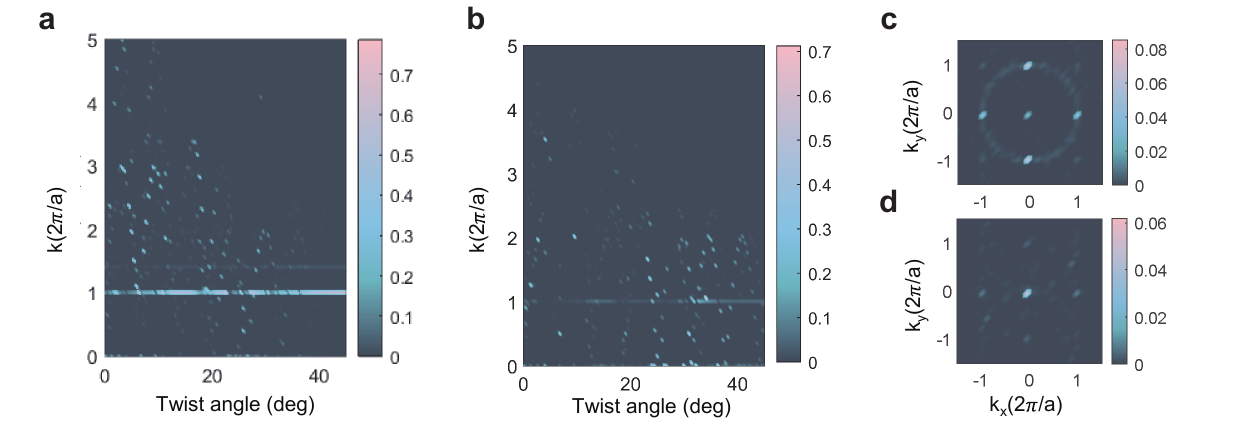}
		\caption{(a) (b)TE and TM component distribution of the eigenmode at $f=0.82c/a$ as a function of twist angle and magnitude of in-plane wave vector. (c) (d) The average TE and TM component distribution of the eigenmode at $f=0.82c/a$ for twist angle from $0^{\circ}$ to $45^{\circ}$.}\label{eigen}
	\end{figure}

	To confirm the legitimacy of the five-layer uniform slab approximation, we conduct an eigenmode analysis in normal incidence case. The eigenvalue in \hyperref[eigen]{Fig.S3(a)} demonstrates a similar angle dependence as the transmission spectra. The valley and summit of transmission spectra exactly correspond to the multi-interference process of the eigenfield. This is straightforward as the constructive interference can maximize the energy transmitted, while the destructive interference can diminish the output. Specifically, we further explore the composition of the twist-angle-independent eigenmodes around $0.82c/a$ as the twist angle changes. In \hyperref[eigen]{Fig.4(a)}, the amplitudes of TE modes with reciprocal lattice vector magnitude equal to $2\pi/a$ are dominating, substantiating the correctness of five-layer uniform slab approximation. Occasionally, some TM mode or other TE mode is dominating causing the breaks in the \hyperref[eigen]{Fig.4(a)}. This can be explained by the influence of twist-angle-dependent resonance. When the different resonances intersect, the resonance modes experience a hybridization process. The breaks correspond to the intersections with twist-angle-dependent resonances, as a matter of fact. On the whole, the dominant Fourier components all reside around the circle $|k|=2\pi/a$ as is shown in the wave vector space in \hyperref[eigen]{Fig.4(c)}. The components with solely kx or ky part are more significant indicating the second slab's influence is weaker compared to the first.      

	The analysis above is appropriate for commensurate scenario as well, except when $\theta=0^{\circ}$. For a twisted bilayer photonic slabs constituted by square lattice slabs, the smallest Pythagorean set is (3,4,5), which has a rather limited effect in the frequency less than $c/a$.

	\subsection{coupled-mode analysis}\label{sec3sec3}
	
\begin{figure}[h]
		\centering
		\includegraphics[width=0.8\textwidth]{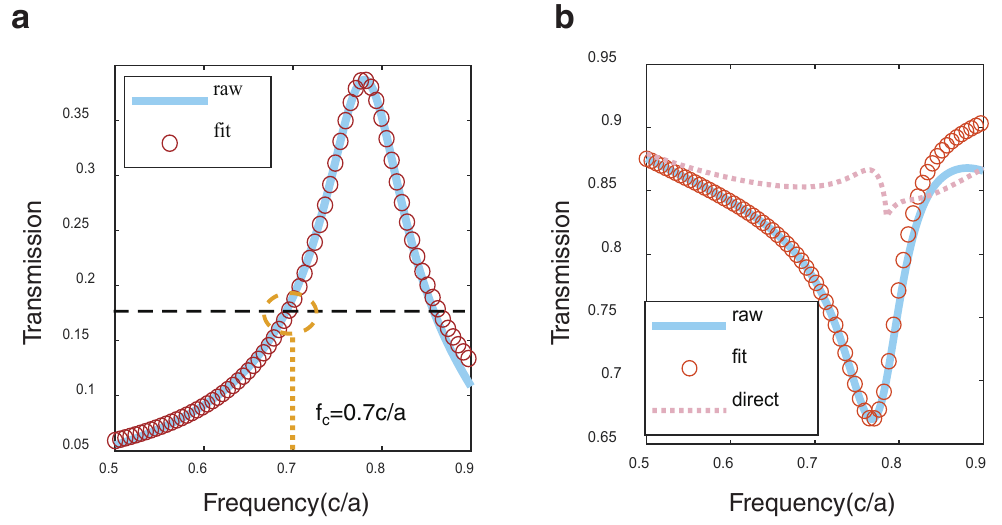}
		\caption{(a) The fitting result for $\left|s^{-}_{\pm1}\right\rangle$. The dashed black line indicates the half of the resonance peak. The lower boundary $f_c=0.7c/a$ of the resonance is marked by the orange circle. (b) The fitting result for $\left|s^{-}\right\rangle$. The fitting parameters value are, $\tau=15.92a/c, \ \tau_{0}=38.82a/c, \ \tau_{1}=9.31a/c$. The fitting trajectory in the right panel deviates from the raw data due to the influence of the photonic crystal mode with higher frequency.}\label{cmt}
	\end{figure}

	Now that we already have a more comprehensive understanding of the origin of the angle-dependent guided resonances, an intriguing question arises: in what frequency range can we find the angle-dependent transmission guided resonance features? While Fig. \hyperref[five-layer]{3(b)} suggests modal existence across all frequencies,  the characteristic resonance signatures are absent below 0.7$c/a$ (Fig. \hyperref[fig1]{1(b)}), where only Fabry-P\'{e}rot cavity-like responses dominate. This is caused by the low coupling strength between the incident light with in-plane wave vector $\boldsymbol{k}_{\text{inc}}$ and the diffracted fields. When the coupling strength is low, the incident field can hardly be scattered into light involving the moir\'{e} wave vector, thus inhibiting resonance excitation. To know more about the transition mechanism between the two transmission phases, the scattering process three described by $\boldsymbol{S}^{(3)}$ is studied by coupled-mode theory\cite{manolatou_coupling_1999,fan_temporal_2003,wonjoo_suh_temporal_2004} with the even photonic crystal slab mode that has the lowest frequency, which is classified by $C_{4v}$ point group into $E$ irreducible representation. In our case, the formalism for CMT is 
	\begin{align}
		&\frac{dA}{dt}=(j\omega_0-\frac{1}{2\tau}-\frac{1}{\tau0}-\frac{2}{\tau_{1}})A+\left\langle \kappa \right|^{*} \left| s_{+}\right \rangle  \\ 		
		&\left|s_{-}\right\rangle=C\left|s_{+}\right\rangle +A\left|D\right\rangle.
	\end{align}\label{cmt1}
	It tells the dynamical equations of the normalized amplitude $A$ of the lowest even $E$ mode at the $\Gamma$ point with resonant frequency $\omega_0$ and lifetime $\frac{\tau}{2}$. The mode can decay into the gap air at a rate of $\frac{1}{\tau_{0}}$ and into the two channels of TE free space eigenmodes with wave vectors $\boldsymbol{G}_{0,1}^{(1)}$ and $\boldsymbol{G}_{0,-1}^{(1)}$ at a rate of $\frac{2}{\tau_{1}}$. The amplitude $A$ accumulates through the coupling of the incoming waves $\left|s_{+}\right\rangle$, with the coupling strength represented by $\left\langle \kappa \right|^{*} $. The outgoing wave $\left|s_{-}\right\rangle$ is formed jointly by the direct pathway $C$ and the indirect pathway $\left|D\right\rangle $, $C=(C_{ij})_{4\times4}$.
	\begin{align}
		&\left|s_{+}\right\rangle=(s^{+},s_0^{+},s_1^{+},s_{-1}^{+})^{\text{T}} \nonumber\\ 		
		&\left|s_{-}\right\rangle=(s^{-},s_0^{-},s_1^{-},s_{-1}^{-})^{\text{T}} \nonumber\\ 
		&\left|\kappa\right\rangle=({\kappa},{\kappa}_0,{\kappa}_1,{\kappa}_{-1})^{\text{T}} \nonumber\\ 
		&\left|D\right\rangle=(d,d_0,d_1,d_{-1})^{\text{T}}\nonumber 
	\end{align}\label{cmt_def}
	The four components stand for the channel in the surrounding air, the gap air, and the TE free space eigenmodes with in-plane wave vector $\boldsymbol{G}_{0,1}^{(1)}$ and in-plane wave vector $\boldsymbol{G}_{0,-1}^{(1)}$(located in the gap air as well) successively. The constants here are not independent, rather they are related by energy conservation and time reversal requirements for the lossless system, yielding $\left|\kappa\right\rangle=\left|D\right\rangle$, $\left|d\right|^{2}=\frac{1}{\tau}$, $\left|d_0\right|^{2}=\frac{2}{\tau_0}$ and $\left|d_{\pm}\right|^{2}=\frac{2}{\tau_1}$ \cite{dean_waves_1984}. The $\omega_0$ and $\tau$ are computed by the standard RCWA\cite{fan_analysis_2002}. The coupling between $s^{+}_{0}$ and $s^{-}_{\pm 1}$ involves only the indirect pathway, since the direct pathway preserves the wave vector, whereas the coupling between $s^{+}_{0}$ and $s^{-}$ involves both the pathways. Assuming the incoming wave state $\left|s_{+}\right\rangle=(0,1,0,0)^{T}$, then
	\begin{align}
		&s^{-}_{\pm1}=\frac{\sqrt{\frac{2}{\tau_0}}\sqrt{\frac{2}{\tau_1}}}{j(\omega-\omega_0)+\frac{1}{2\tau}+\frac{1}{\tau_0}+\frac{2}{\tau_1}}\\ 		
		&s^{-}=C_{1,2}+\frac{\sqrt{\frac{1}{\tau}}\sqrt{\frac{2}{\tau_0}}}{j(\omega-\omega_0)+\frac{1}{2\tau}+\frac{1}{\tau_0}+\frac{2}{\tau_1}}.
	\end{align}\label{cmt_res}
	The phase factor in Eq.\hyperref[cmt_res]{(6) and (7)} is disposed by appropriate choice of the reference plane of the channels. For $f \textless c/a$, $s^{-}_{1}$ and $s^{-}_{-1}$, the indices of coupling strength, are solely determined by the resonance, whereas $s^{-}$ and $s^{-}_{0}$ remain finite outside the resonance regime. The resonance width broadens due to the decay rate of the expanded channels introduced by the modified eigenmodes in contrast to the original linewidth calculated by the usual free space eigenmode set. The lower boundary of the resonance coincides with the critical frequency $f_c=0.7c/a$ that demarcates the angle-dependent resonance and Fabry-P\'{e}rot resonance regimes. Below $f_c$, twist-angle-independent transmission spectra vanish, indicating the absence of guide mode coupling. It should be noted, however, that there can still be a characteristic transmission spectra above this resonance range. This is because photonic crystal slab modes with higher frequency, which are a combination of modes with larger in-plane wave vectors can be stimulated and scatter the incident field into resonance. In addition, we fit the $\left|s^{-}_{\pm1}\right\rangle$ and $\left|s^{-}\right\rangle$ to determine the values of $\tau_{0}$ and $\tau_{1}$ in order to examine the legitimacy of applying coupled-mode theory. A slight center frequency drift, owing to the expanded channels, is observed and both results fit well with the computed data. The fitting trajectory in the right panels of Fig.5\hyperref[cmt]{(a)} and Fig.5\hyperref[cmt]{(b)} deviates from the raw data due to the influence of the modes with higher frequency. Therefore, we can conclude that the characteristic angle-dependent transmission spectra emerge from the coupling of the original channels and the expanded channels of the modified eigenmodes, with the lowest-frequency single slab's resonance's lower boundary ($f_c=0.7c/a$) indicating the critical transition between angle-dependent resonance and Fabry-P\'{e}rot resonance regimes.
	
	\section{Discussion}\label{discussion}
	In this paper, we have first propose, to the best of our knowledge, the modified RCWA, where the evanescent bases are substituted by bases with non-zero energy flux density to appropriately describe the multiple scattering process between the two slabs. With this new approach, we demonstrate the calculation of the eigenmodes and the transition frequency between the angle-dependent resonance and Fabry-p\'{e}rot resonance, which have not ever been realized before. A five-layer uniform slab approximation is also proposed to predict the resonance of the twisted bilayer system with an accuracy around 0.04a/c. However, the presence of the moir\'{e} pattern induce a split in the five-layer resonance modes, which shows great potential for engineering the band structure of photonic crystal. Note that this approximation can be easily extrapolated to twisted multi-layer structure and is straightforward for optical engineers to implement. Our study explores the resonance behavior of the twisted bilayer structure and serves as a foundation for the next generation LiDAR optical steering module, laser and reconfigurable devices. 
	

	\bibliography{main,RCWA,CMT,pcsel,lidar}%
	
		
		
		
		
	
\end{document}


\maketitle
	\section{The reason to use new eigenmodes}
\begin{figure}[h]
		\centering
		\begin{minipage}{1\linewidth}
			\centering
			\includegraphics[width=0.5\linewidth,height=5.5cm]{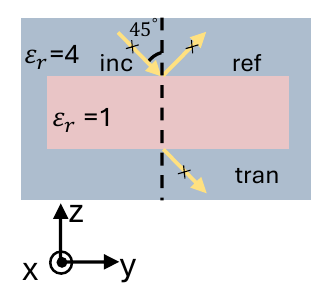}
			
		\end{minipage}
		\caption{  Light obliquely incident on the air slab from silicon nitride ($\varepsilon_r=4$) at an angle of $45^{\circ}$, and the field within the slab is represented by the linear combination of evanescent waves. The reflection coefficient for the evanescent wave in the air slab is $(2\sqrt{2}\mathrm{i}-1)/3$.}\label{figs1}
	\end{figure}

	To illustrate the necessity of modification, we consider a simple example of a TE plane wave obliquely incident into an air slab from silicon nitride ($\varepsilon_r=4$) at an angle of $45^{\circ}$. The field in the air slab is represented by the linear combination of evanescent waves as shown in Fig.\hyperref[figs1]{s1}. The reflection coefficient of the evanescent wave in the air slab can be derived from electromagnetic boundary conditions, yielding $(2\sqrt{2}\mathrm{i}-1)/3$. If one were to use the reflection coefficient to perform a multiple scattering process between the two slabs to calculate the final field distribution, it would diverge. Because, after the scattering, the reflected evanescent wave's amplitude stays invariant in magnitude, while for a general Fabry-P\'{e}rot cavity, the reflected wave's amplitude decreases exponentially. This fundamental distinction arises because evanescent waves exhibit zero Poynting vector component along their evanescent decay direction. The energy conservation law enforces the attenuation in the scattering process for waves with a non-zero Poynting vector along the propagating direction. Combining counter-propagating evanescent components to form a new set of eigenfunctions can resolve this divergence (more detailed information is in Supplemental Material, Sec. 1).
	\section{The new eigenmodes}

	Before exploring the new eigenmodes, it is instructive to revisit the fundamental basis of free-space electromagnetic eigenmodes. For electromagnetic waves characterized by an in-plane wave vector $\boldsymbol{k} = (k_x,k_y)^{\mathsf{T}}$, the system supports two mutually orthogonal eigenmodes: transverse electric (TE) and transverse magnetic (TM) polarizations. These eigenstates can be formulated through their in-plane electric field components:
	
	\begin{itemize}
		\item TE mode: 
		$\displaystyle \frac{1}{\sqrt{\lvert k_z\rvert(k_x^2 + k_y^2)}}\begin{pmatrix} -k_y \\ k_x \end{pmatrix}$
		
		\item TM mode:
		$\displaystyle \frac{\sqrt{\lvert k_z \rvert}}{k\sqrt{k_x^2 + k_y^2}} \begin{pmatrix} k_x \\ k_y \end{pmatrix}$
	\end{itemize}
	where $k = \lvert \boldsymbol{k}\rvert$ represents the wave number magnitude. The longitudinal field component $E_z$ can be derived through the application of the Gauss' flux theorem. Notably, these mode profiles are normalized with respect to the $z$-component of the time-averaged Poynting vector, ensuring unit power flux in the propagation direction.
	
	As expected, the new eigenmodes should satisfy two fundamental requirements: (i) maintain non-zero z-component of the Poynting vector for directional energy propagation(along z axis), and (ii) preserve the time-reversal symmetry requirement for the scattering matrix \cite{dean_waves_1984}. These constraints lead to modified eigenmode construction through superposition of counter-propagating evanescent waves. The complete parameterization for these hybrid modes is tabulated in \hyperref[tables1]{Table S1}, where $a=e^{\lvert k_z \rvert h}$, $E_{te}$ and $E_{tm}$ are the original TE, TM modes respectively. Notably, the backward-propagating counterparts can be generated through complex conjugation operations, while there is no need to modify the traveling modes. This construction ensures a straightforward and self-consistent description for multiple scattering processes between the two dielectric slabs.

	\begin{table}[htbp]
		\centering
		\begin{tabular}{|c|c|c|c|c|}
			\hline
			\rowcolor[rgb]{ .914,  .773,  .776} scattering process & $TE_+$ & $TE_-$ & $TM_+$ & $TM_-$ \bigstrut\\
			\hline
			$S^1$ & $(i\frac{E_{te}^{-}}{\sqrt{2a}}+\sqrt{a}\frac{E_{te}^{+}}{\sqrt{2}})$  & $(-i\frac{E_{te}^{-}}{\sqrt{2a}}+\sqrt{a}\frac{E_{te}^{+}}{\sqrt{2}})$  & $(\frac{E_{tm}^{-}}{\sqrt{2a}}+i\sqrt{a}\frac{E_{tm}^{+}}{\sqrt{2}})$ & $(\frac{E_{tm}^{-}}{\sqrt{2a}}-i\sqrt{a}\frac{E_{tm}^{+}}{\sqrt{2}})$ \bigstrut\\
			\hline
			\rowcolor[rgb]{ .698,  .843,  .922} $S^2$ & $(i\sqrt{a}\frac{E_{te}^{-}}{\sqrt{2}}+\frac{E_{te}^{+}}{\sqrt{2a}})$   & $(-i\sqrt{a}\frac{E_{te}^{-}}{\sqrt{2}}+\frac{E_{te}^{+}}{\sqrt{2a}})$   & $(\sqrt{a}\frac{E_{tm}^{-}}{\sqrt{2}}+i\frac{E_{tm}^{+}}{\sqrt{2a}})$ & $(\sqrt{a}\frac{E_{tm}^{-}}{\sqrt{2}}-i\frac{E_{tm}^{+}}{\sqrt{2a}})$ \bigstrut\\
			\hline
			$S^3$ & $(i\frac{E_{te}^{-}}{\sqrt{2a}}+\sqrt{a}\frac{E_{te}^{+}}{\sqrt{2}})$  & $(-i\frac{E_{te}^{-}}{\sqrt{2a}}+\sqrt{a}\frac{E_{te}^{+}}{\sqrt{2}})$  & $(\frac{E_{tm}^{-}}{\sqrt{2a}}+i\sqrt{a}\frac{E_{tm}^{+}}{\sqrt{2}})$ & $(\frac{E_{tm}^{-}}{\sqrt{2a}}-i\sqrt{a}\frac{E_{tm}^{+}}{\sqrt{2}})$ \bigstrut\\
			\hline
		\end{tabular}%
		\label{tab:addlabel}%
		\caption{Alternative forward and backward eigenmodes for evanescent modes for each scattering matrix.}
	\end{table}%
	
	\section{Constructing scattering matrix using RCWA}
	In a periodic structure, the dielectric constant can be expand into Fourier series, while the electric filed and magnetic field are represented by Floquet-Fourier series\cite{li_new_1997}. 
	
	\begin{align}
		&\epsilon(x,y,z)=\sum_{m,n} \varepsilon_{m,n}(z) e^{i\frac{2\pi}{d_1}mx+i\frac{2\pi}{d_2}ny}  \\ 		
		&\phi^i=\sum_{m,n} \phi_{m,n}^i(z) e^{i\frac{2\pi}{d_1}mx+i\frac{2\pi}{d_2}ny}e^{i \boldsymbol{k}_{\text{inc}} \cdot\boldsymbol{r} } \qquad i=x,y,z.
	\end{align}\label{expand}
	The $\phi^i,\enspace i=x,y,z$ stand for three components of the electric field or magnetic field($\sqrt{\frac{\mu_0}{\varepsilon_0}}\boldsymbol{H}$) respectively, $\boldsymbol{k}_{inc}$ is the in-plane wave vector of the incident field, $d_1$, $d_2$ are the period along x and y axis respectively and $\mu$ is the relative permeability. Denote $\alpha_{mn}=\frac{2\pi}{d_1}m/k$, $ \beta_{mn}=\frac{2\pi}{d_2}n/k$ and $\widetilde{z}=kz$, $k$ is the magnitude of the wave vector in free space. In the case of single-layer $\varepsilon$ is independent of z. When expressed through the aforementioned expansions, Maxwell's equations assume the following representation
	\begin{align}
		&\beta_{mn}E^z_{mn}+i\partial_{\widetilde{z}} E^y_{mn}=\mu H^x_{mn}\\ 		
		&-i\partial_{\widetilde{z}} E^x_{mn}-\alpha_{mn}E^z_{mn}=\mu H^y_{mn}\\ 
		&\alpha_{mn}E^y_{mn}-\beta_{mn} E^x_{mn}=\mu H^z_{mn}\\ 
		&\beta_{mn}H^z_{mn}+i\partial_{\widetilde{z}} H^y_{mn}=-\sum_{i,j}E^x_{i,j}\varepsilon_{m-i,n-j}\\ 
		&-i\partial_{\widetilde{z}} E^x_{mn}-\alpha_{mn}E^z_{mn}=-\sum_{i,j}E^y_{i,j}\varepsilon_{m-i,n-j}\\ 
		&\alpha_{mn}E^y_{mn}-\beta_{mn} E^x_{mn}=-\sum_{i,j}E^z_{i,j}\varepsilon_{m-i,n-j}
	\end{align}\label{expandm}
	Note that, we assume a ${\rm exp}(-i\omega t)$ time dependence and the material being nonmagnetic. Then we transform the equations above into matrix format. 
	\begin{align}
		&\boldsymbol{\beta}\boldsymbol{E^z}+i\partial_{\widetilde{z}} \boldsymbol{E^y}=\mu \boldsymbol{H^x}\\ 		
		&-i\partial_{\widetilde{z}}\boldsymbol{E^x}-\boldsymbol{\alpha}\boldsymbol{E^z}=\mu \boldsymbol{H^y}\\ 
		&\boldsymbol{\alpha}\boldsymbol{E^y}-\boldsymbol{\beta} \boldsymbol{E^x}=\mu \boldsymbol{H^z}\\ 
		&\boldsymbol{\beta}\boldsymbol{H^z}+i\partial_{\widetilde{z}} \boldsymbol{H^y}=-\boldsymbol{\varepsilon}\boldsymbol{E^x}\\ 
		&-i\partial_{\widetilde{z}} \boldsymbol{H^x}-\boldsymbol{\alpha}\boldsymbol{H^z}=-\boldsymbol{\varepsilon}\boldsymbol{E^y}\\ 
		&\boldsymbol{\alpha}\boldsymbol{H^y}-\boldsymbol{\beta} \boldsymbol{H^x}=-\boldsymbol{\varepsilon}\boldsymbol{E^z}
	\end{align}\label{expandmatrix}
	$\boldsymbol{\varepsilon}$ is the convolution matrix of the dielectric constant\cite{li_new_1997,lou_theory_2021}. $\boldsymbol{\Phi^i}= \begin{pmatrix} 
		\phi_{m_1,n_1}^i \\
		\phi_{m_2,n_2}^i \\
		\vdots \\
		\phi_{m_M,n_M}^i 
	\end{pmatrix}$, $i=x,y,z$ stand for the vector of corresponding physical element and $\boldsymbol{\beta}$, $\boldsymbol{\alpha}$ is the corresponding diagonal matrix. $\boldsymbol{\Phi^i}$ corresponds to a permutation of reciprocal lattice vectors. M indicates the number of bases used in the calculation. The z component field can be eliminated by equations \hyperref[expandmatrix]{(S11), (S14)} and the final equation is 
	\begin{align}
		-i\frac{\partial}{\partial_{\widetilde{z}}}\begin{pmatrix}
			E^x \\ E^y
		\end{pmatrix}
		=F\begin{pmatrix}
			H^x \\ H^y
		\end{pmatrix}
		\\
		-i\frac{\partial}{\partial_{\widetilde{z}}}\begin{pmatrix}
			H^x \\ H^y
		\end{pmatrix}
		=G\begin{pmatrix}
			E^x \\ E^y
		\end{pmatrix}
	\end{align} \label{partial}
	\begin{align}
		F=\begin{pmatrix}
			\boldsymbol{\alpha} \boldsymbol{\varepsilon}^{-1} \boldsymbol{\beta} & \mu -\boldsymbol{\alpha} \boldsymbol{\varepsilon}^{-1} \boldsymbol{\alpha} \\ \boldsymbol{\beta} \boldsymbol{\varepsilon}^{-1} \boldsymbol{\beta}-\mu & -\boldsymbol{\beta} \boldsymbol{\varepsilon}^{-1}\boldsymbol{\alpha}
		\end{pmatrix} ,\enspace  G=\begin{pmatrix}
			-\boldsymbol{\alpha\beta}/\mu& \boldsymbol{\alpha\alpha}/\mu-\boldsymbol{\varepsilon} \\ \boldsymbol{\varepsilon}-\boldsymbol{\beta\beta}/\mu & \boldsymbol{\beta\alpha}/\mu
		\end{pmatrix}
	\end{align}
	Assuming the field in equations \hyperref[partial]{(S15), (S16)} is proportional to ${\rm exp}(i\gamma z)$, then we can arrive at an eigenvalue problem by combining the two equations.
	\begin{align}
		(FG-\gamma^2)\begin{pmatrix}
			E^x \\ E^y
		\end{pmatrix}
		=\boldsymbol{0}
	\end{align} \label{partial1} 
	
	By solving the eigenvalue problem of matrix FG, can we get the eigenfunctions in the single-layer periodic structure, denoted by $\boldsymbol{W}$ and $\boldsymbol{V}$, the eigenmode matrix for electric field and magnetic field respectively. $\boldsymbol{V}$ is related to $\boldsymbol{W}$ by
	\begin{align}
		\boldsymbol{\Gamma}\boldsymbol{V}=G\boldsymbol{W}
	\end{align} \label{relation}
	
	Here $\boldsymbol{\Gamma}$ denotes the diagonal matrix for the eigenvalue $\gamma$. The modified free space eigenmodes upward and downward are denoted as $\boldsymbol{W}_1,\enspace \boldsymbol{W}_2$ for electric field and $\boldsymbol{V}_1, \enspace \boldsymbol{V}_2$ for magnetic field. $S^{(i)}(\boldsymbol{k}_{\text{inc}})$ is defined by the equation below.
	\begin{align}
		\begin{pmatrix}
			\boldsymbol{C_1^-} \\ \boldsymbol{C_2^+} 
		\end{pmatrix}=
		S^{(i)}(\boldsymbol{k}_{\text{inc}}) \begin{pmatrix}
			\boldsymbol{C_1^+} \\ \boldsymbol{C_2^-} 
		\end{pmatrix} 
	\end{align}
	The scattering matrix $S^{(i)}(\boldsymbol{k}_{\text{inc}})$ can be extracted by the boundary condition equations below\cite{rumpf_improved_2011}.
	\begin{align}
		\begin{pmatrix}
			\boldsymbol{W_1} & \boldsymbol{W_2} \\ \boldsymbol{V_1} & \boldsymbol{V_2}
		\end{pmatrix} 
		\begin{pmatrix}
			\boldsymbol{C_1^+}  \\ \boldsymbol{C_1^-} 
		\end{pmatrix}
		=\begin{pmatrix}
			\boldsymbol{W}& \boldsymbol{W} \\ \boldsymbol{V}& \boldsymbol{-V}
		\end{pmatrix}
		\begin{pmatrix}
			\boldsymbol{C^+} \\ \boldsymbol{C^-} 
		\end{pmatrix} \\
		\begin{pmatrix}
			\boldsymbol{W} & \boldsymbol{W} \\ \boldsymbol{V} & \boldsymbol{-V}
		\end{pmatrix} 
		\begin{pmatrix}
			e^{i\boldsymbol{\Gamma}h}& 0 \\ 0& e^{-i\boldsymbol{\Gamma}h}
		\end{pmatrix}
		\begin{pmatrix}
			\boldsymbol{C^+}  \\ \boldsymbol{C^-} 
		\end{pmatrix}
		=\begin{pmatrix}
			\boldsymbol{W_1^{'}}& \boldsymbol{W_2^{'}} \\ \boldsymbol{V_1^{'}} & \boldsymbol{V_2^{'}}
		\end{pmatrix}
		\begin{pmatrix}
			\boldsymbol{C_2^+} \\ \boldsymbol{C_2^-} 
		\end{pmatrix} &&
	\end{align}

	\begin{figure}[h]
		\centering
		\begin{minipage}{1\linewidth}
			\centering
			\includegraphics[width=1\linewidth,height=5.5cm]{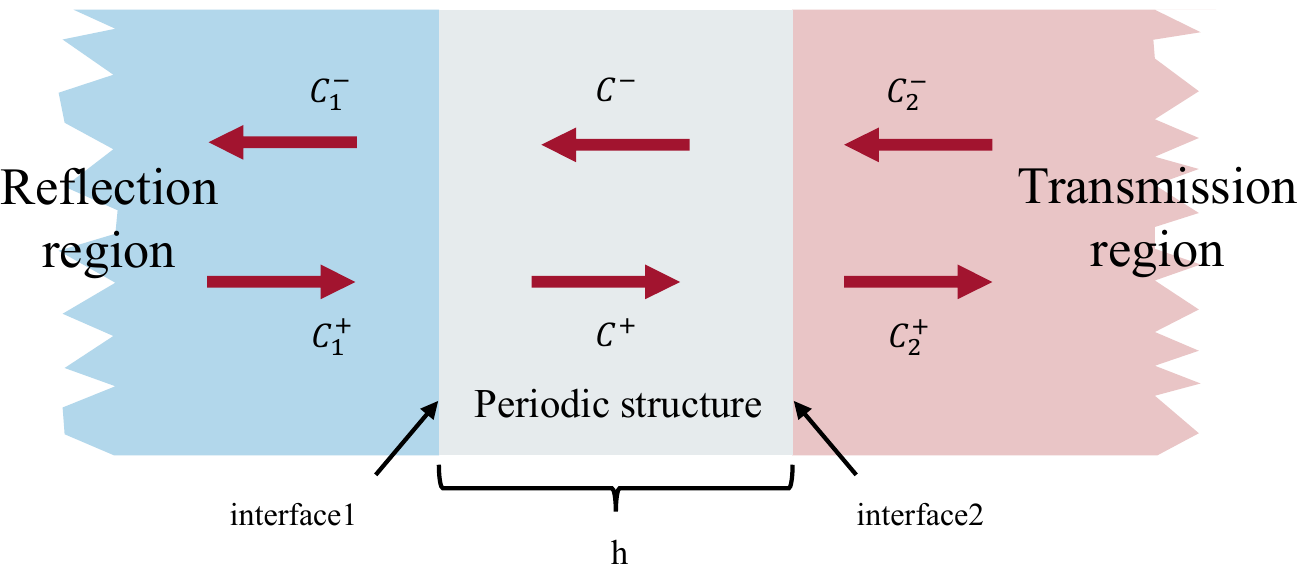}
			
		\end{minipage}
		\caption{ Physical process of scattering involved with one periodic structure. The amplitude of the corresponding forward and backward eigenmodes in each section is connected by the boundary condition that the tangential electric field and magnetic field are equal at the two interfaces.}\label{fig1}
	\end{figure}
	
	The choice for the free space eigenmodes $W_1$, $W_2$, $V_1$, and $V_2$ depends on the boundary condition. It is appropriate to use original eigenmodes outside the two slabs, while the modified free space eigenmodes should only be used in the air gap between the two slabs due to the implicit nature boundary condition that the field at the infinity must converge. By solving the equations above, one can obtain the corresponding single-layer scattering matrix $S^{(i)}(\boldsymbol{k}_{\text{inc}})$. 
	
	Denoting $\mathcal{L}_1\equiv\{2\pi/d_1m\hat{i}+2\pi/d_2n\hat{j} \textbar m,n\in\mathbb{Z}\}\equiv\{\boldsymbol{G}_{m,n}^{(1)}\textbar m,n\in\mathbb{Z}\}$, $\mathcal{L}_1\equiv\{2\pi/d_1m\hat{i^{'}}+2\pi/d_2n\hat{j^{'}} \textbar m,n\in\mathbb{Z}\}\equiv\{\boldsymbol{G}_{m,n}^{(2)}\textbar m,n\in\mathbb{Z}\}$
	
	Denoting $\boldsymbol{G}_{m,n}^{(1)}\equiv2\pi/d_1m\hat{i}+2\pi/d_2n\hat{j}$ and $\boldsymbol{G}_{m,n}^{(2)}\equiv2\pi/d_1m\hat{i^{'}}+2\pi/d_2n\hat{j^{'}}$, with $\hat{i}$ and $\hat{j}$ denoting the unit vectors along the positive x-axis and y-axis respectively, $\hat{i^{'}}=cos(\theta)\hat{i}+sin(\theta)\hat{j}$ and $\hat{j^{'}}=-sin(\theta)\hat{i}+cos(\theta)\hat{j}$, where $\theta$ is the twist angle.
	
	Denote $\mathcal{L}=\mathcal{L}_1\oplus\mathcal{L}_2$. $\mathcal{L}$ represents the direct sum of the reciprocal spaces of slab 1 and slab 2.
	
	To calculate the scattering matrix for the twisted bilayer system, some simple changes are necessary. Standard RCWA constructs scattering matrices involved with solely modes with in-plane wave vector of  $\boldsymbol{k}_n=\boldsymbol{G}_{m_1,n_1}^{(i)}+\boldsymbol{k}_{\text{inc}}$, $i=1,2$. To cover the modes labeled by both slabs' reciprocal spaces, one can sweep $\boldsymbol{k}_{\text{inc}}$ across the reciprocal lattice space of either slab and synthesize the scattering matrices into a comprehensive one\cite{lou_theory_2021}. The arrangement of the scattering matrices depends on the permutation of the reciprocal lattice vectors of both the slabs. 
	
	Denote $\boldsymbol{\pi}_1$ as the permutation for slab 1 and $\boldsymbol{\pi}_2$ for slab 2. $\boldsymbol{\pi}$ represents the permutation for the two slabs. A typical relation among $\boldsymbol{\pi}_1$, $\boldsymbol{\pi}_2$ and $\boldsymbol{\pi}$ is $\boldsymbol{\pi}=[\boldsymbol{\pi}_1+\boldsymbol{\pi}_2(1),\boldsymbol{\pi}_1+\boldsymbol{\pi}_2(2),\boldsymbol{\pi}_1+\boldsymbol{\pi}_2(3), \cdots]$. In this case, the synthesis processes are exemplified below. 
	\begin{figure}[h]
		\centering
		\begin{minipage}{1\linewidth}
			\centering
			\includegraphics[width=1\linewidth]{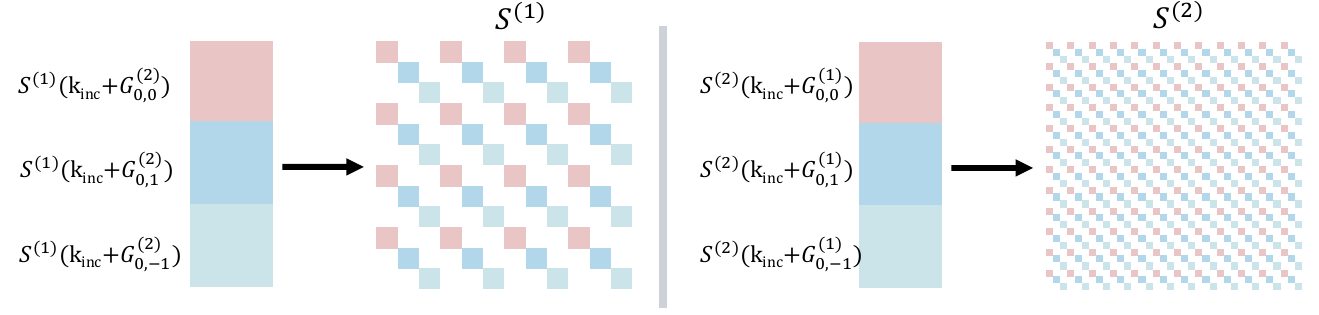}		
		\end{minipage}
		\caption{ The synthesis processes for the scattering matrices of slab 1 and slab 2. The left and right panel represent the synthesis process for slab 1 and the slab 2 respectively. In this case, only three reciprocal lattice vectors are considered. }\label{zonghe}
	\end{figure}
	
	\section{Time-domain spectroscopy technique}
\begin{figure}[h]
		\centering
		\begin{minipage}{1\linewidth}
			\centering
			\includegraphics[width=1\linewidth,height=5.5cm]{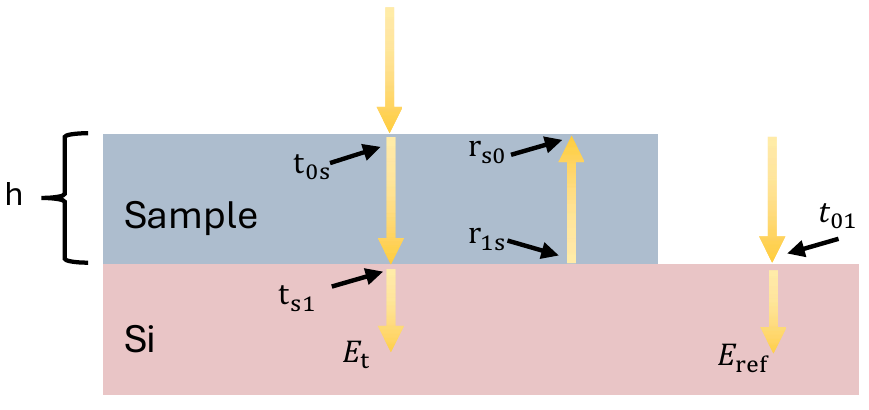}
			
		\end{minipage}
		\caption{The schematic for theory of the Time-domain spectroscopy technique. The signal is received in the silicon substrate.  }\label{tds}
	\end{figure}

	Time-domain spectroscopy(TDS) technique compute the dielectric constant of the system by measuring the transmission or reflection spectra of the sample. By utilizing the ratio of the transmission or reflection spectro with and without the sample, a complex equation about the dielectric constant is obtained. One can employ numerical technique to solve this complex equation \hyperref[ratio]{S20}. Figure \hyperref[tds]{S2} schematically illustrates the principles of TDS operation

	In monochromatic light scenarios, the ratio can be directly determined through reflection and transmission coefficients. However, when analyzing time-domain signals, a distinct approach is required: the temporal waveform of the transmitted signal must first be converted to the frequency domain via Fourier transformation. This spectral decomposition enables the establishment of a frequency-dependent relationship between the refractive index n and angular frequency $\omega$. By systematically solving the characteristic electromagnetic wave equation in Eq. \hyperref[ratio]{S20} across the relevant frequency spectrum, one can derive the dispersion relation n($\omega$). 
	
	\begin{equation}
		\frac{E_t}{E_{ref}}=\frac{t_{0s}t_{s1}e^{ik_0(n-1)h}}{t_{0s}(1-r_{1s}r_{s0}e^{2ik_0nh})}
	\end{equation}\label{ratio}
	
	The result is shown below. 
	
	\section{Accuracy of Five-layer Uniform Slab Approximation}
	To examine the accuracy of the appoximation method, we calculate the twist bilayer structure for a range of parameter and compare the approximate angle-independent resonance frequency to the corresponding result of RCWA. We sweep three times: r from 0.1 to 0.45a, d=0.3a, h=0.2a; r=0.3a d from 0.1 to 0.4a, h=0.2a; r=0.3a d=0.3a h from 0.1 to 0.35. As a result, the approximation attain an accuracy around 0.04c/a.     
	
	\section{Split of Resonance}
	The twist bilayer system conform a $D_4$ symmetry, which is more clear if we specify the coordinate system as Fig.S5. $D_4=\{E,C_4,C_4^{-1},C_4^2,\sigma_x,\sigma_y,\sigma_{d1},\sigma_{d2} \}$. In a perturbation perspective, the field can be expressed as $\Phi=\sum_{\phi_{i}\in V}a_i\phi_{i}$. To explain the split of resonance at $f=0.82c/a$, we use $\phi_{i}$ as bases to express the $D_4$ group. Below, we show the two set of doubly degenerate E modes and the corresponding representation of some of the element in $D_4$ group. 
	\begin{align}
		&E^{(1)}_1:(1,-i,1,i)^{\text{T}} \\ 		
		&E^{(1)}_2:(i,1,-i,1)^{\text{T}} \\ 
		&E^{(2)}_1:(1,-i,-1,-i)^{\text{T}} \\ 
		&E^{(2)}_2:(i,1,i,-1)^{\text{T}} 
	\end{align}\label{E}
	
	\begin{align}
		C_4=
		\begin{pmatrix}
			0 & -1 &0 & 0\\ 
			1 & 0 &0 & 0\\ 
			0 & 0 &0 & -1\\ 
			0 & 0 &1 & 0\\ 
		\end{pmatrix} 
	\end{align}\label{c4}
	
	\begin{align}
		C_4=
		\begin{pmatrix}
			1 & 0 &0 & 0\\ 
			0 & -1 &0 & 0\\ 
			0 & 0 &-1 & 0\\ 
			0 & 0 &0 & 1\\ 
		\end{pmatrix} 
	\end{align}\label{xigema}
	The four component in \hyperref[E]{S24} to \hyperref[E]{S27} refer to the basis $\phi_{i}$ with zero in-plane wave vector polarized along the four basic reciprocal lattice vector of the two slabs respectively.
	
	\bibliography{main}